\documentclass[twocolumn,english,floatfix,aps,prd,showpacs,amsmath,amssymb]{revtex4-1}
\usepackage[T1]{fontenc}
\usepackage[latin9]{inputenc}
\usepackage{color}
\usepackage{babel}
\usepackage{amstext}
\usepackage{amssymb}
\usepackage{graphicx}

\begin{document}

\title{Dynamical Casimir effect with $\delta-\delta^{\prime}$ mirrors}

\author{Jeferson Danilo L. Silva}
\email{danilo.fisic@gmail.com}
\email{jeferson.silva@icen.ufpa.br}
\affiliation{Faculdade de Física, Universidade Federal do Pará, Belém, Pará, Brazil }

\author{Alessandra N. Braga}
\email{alessandrabg@ufpa.br}
\affiliation{Faculdade de Física, Universidade Federal do Pará, Belém, Pará, Brazil }

\author{Danilo T. Alves}
\email{danilo@ufpa.br}
\affiliation{Faculdade de Física, Universidade Federal do Pará, Belém, Pará, Brazil }

\date{\today}

\begin{abstract}
We calculate the spectrum and the total rate of created particles for a real massless scalar field in $1+1$ dimensions, in the presence of a partially transparent moving mirror simulated by a Dirac $\delta-\delta^{\prime}$ point interaction. 
We show that, for this model, a partially reflecting mirror can produce a larger number of particles in comparison with a perfect one.
In the limit of a perfect mirror, our formulas recover those found in the literature for the particle creation by a moving mirror with a Robin boundary condition. 
\end{abstract}


\maketitle
%%%%%%%%%%%%%%%%%%%%%%%%%%%%%%%%%%%%%%%%%%%%%%%%%%%%%%%%%%%%%%%%%%%%%
\section{Introduction}
\label{intro}

Real particles can be generated from the vacuum when a quantized field
is submitted to time-dependent boundary conditions. This phenomenon
is usually called the dynamical Casimir effect (DCE). It was first
investigated in the 1970's decade in theoretical papers by Moore
\cite{Moore-1970}, DeWitt \cite{DeWitt-1975}, Fulling and Davies
\cite{Fulling-Davies-1976-1977,Davies-Fulling-1977}, and Candelas
and Deutsch \cite{Candelas-1977}. Nowadays, the available literature
on the DCE is quite wide (see Refs. \cite{Dodonov-2009-2010,Dalvit-MaiaNeto-Mazzitelli-2011}
for a detailed review). In 2011, Wilson \emph{et al.} \cite{Johansson-Nature-2011}
observed experimentally the DCE by the first time, in the context of circuit Quantum Electrodynamics. Namely,
a time-dependent magnetic flux is applied in a coplanar waveguide
(transmission line) with a superconducting quantum interference device
(SQUID) at one of the extremities, changing the inductance of the
SQUID, and thus yielding a time-dependent boundary condition \cite{Johansson-Nature-2011,Johansson-2009}.
Another observation of the DCE was announced by Lähteenmäki \emph{et
al}. \cite{Lahteenmaki-2013}. Some other experimental proposals aiming
the observation of the DCE can be found in Ref. \cite{Proposals-observation-DCE}.

During the first two decades after the paper by Moore \cite{Moore-1970},
calculations on the DCE were usually done with perfectly reflecting
mirrors. In this context, expressions for the force acting on the mirror and the radiated
energy have been derived in Refs. \cite{DeWitt-1975,Fulling-Davies-1976-1977,Davies-Fulling-1977,Candelas-1977,Ford-Vilenkin-1982}.
On the other hand, as Moore has pointed out in Ref. \cite{Moore-1970}, real mirrors
do not behave as perfectly reflecting at all and, moreover, the formula
for the radiated energy by a perfect mirror, obtained in Ref. \cite{Fulling-Davies-1976-1977},
exhibits an inconsistency: the renormalized energy can be negative
when the mirror starts moving, and thus it can not be associated with
the energy of the created particles \cite{Fulling-Davies-1976-1977,Haro-Elizalde-2006}.
Haro and Elizalde \cite{Haro-Elizalde-2006} showed that when a partially
reflecting mirror is considered, this inconsistency can be avoided,
and the radiated energy is always positive.

The DCE with partially reflecting mirrors has been investigated by 
several authors (see, for instance, \cite{Haro-Elizalde-2006,Jaekel-Reynaud-1992,Eberlein-1993,
Lambrecht-Jaekel-Reynaud-1996,Lambrecht-Jaekel-Reynaud-1998,Obadia-Parentani-2001,
Barton-Calogeracos-1995-I,Nicolaevici-2001,Nicolaevici-2009}).
Dirac $\delta$ potentials for modeling partially reflecting moving mirrors were considered
by Barton and Calogeracos \cite{Barton-Calogeracos-1995-I}.
These authors investigated 
the radiation reaction force for a $\delta$ moving mirror 
in the nonrelativistic regime. When the mirror is at rest,
the model is given explicitly by (hereafter
$c=\hbar=1$) \cite{Barton-Calogeracos-1995-I} 
\begin{equation}
\mathcal{L}=(1/2)[(\partial_{t}\phi)^{2}-(\partial_{x}\phi)^{2}]-\mu\delta(x)\phi^{2}(t,x),
\label{eq:Barton}
\end{equation}
where $\mu$ is related to the plasma frequency, since 
this model is a good approximation
for the interaction between the electromagnetic field and a plasma thin
sheet \cite{Barton-Calogeracos-1995-I}. The transmission
and reflection coefficients associated to (\ref{eq:Barton}) are respectively
\begin{equation}
s_{\pm}(\omega)=\frac{\omega}{\omega+i\mu}\quad\text{and}\quad r_{\pm}(\omega)=\frac{-i\mu}{\omega+i\mu}.
\label{eq:s-and-r-barton}
\end{equation}
The labels ``$+$'' and ``$-$'' represent the scattering to the
right and to the left of the mirror respectively [in this case $s_{+}(\omega)=s_{-}(\omega)$
and $r_{+}(\omega)=r_{-}(\omega)$]. The transparency of the mirror can be controlled
by tuning $\mu$. Particularly, the limit $\mu\rightarrow\infty$
leads straightforwardly to the well-known Dirichlet boundary condition
in both sides of the mirror, namely (see the Appendix for details)
\begin{equation}
\phi_{+}(t,0^{+})=0\quad\text{and}\quad\phi_{-}(t,0^{-})=0,\quad(\mu\rightarrow\infty),
\label{limite-dirichlet-barton}
\end{equation}
where $\phi_{+}(t,x)$ and $\phi_{-}(t,x)$ represent the field in
the right and left sides of the mirror respectively. The generalization
to a relativistic moving mirror was done in Ref. \cite{Nicolaevici-2001}.
The model (\ref{eq:Barton}) was also considered in the investigation
of the static Casimir effect \cite{Castaneda-Guilarte-2013} and DCE (in connection with decoherence \cite{MaiaNeto-Dalvit-2000} and Hawking
radiation \cite{Nicolaevici-2009}). 
Beyond the scalar field, the consideration of partially reflecting mirrors by means of terms in the Lagrangian involving 
$\delta$ functions, as in Eq. (\ref{eq:Barton}), have been done by Barone and Barone who proposed 
an enlarged gauge-invariant Maxwell Lagrangian describing the field in the presence of a $\delta$ mirror,
investigating the interaction between a static charge and a $\delta$ mirror \cite{Barone-2014}, and also the static Casimir effect between two $\delta$ mirrors \cite{Barone-2014-2}.
Parashar, Milton, Shajesh and Schaden \cite{Parashar-Milton-Shajesh-Shaden-2012}, in a different approach, considered the electric and magnetic properties of an infinitesimally thin mirror by means of the electric permittivity and magnetic permeability described in terms of $\delta$ functions,
deriving the boundary conditions for the electromagnetic field on the mirror and investigating the Casimir-Polder interaction between an atom and the mirror.

The use of $\delta-\delta^{\prime}$ potentials ($\delta^{\prime}$
is the derivative of the Dirac $\delta$) for simulating partially
reflecting mirrors, in the context of the static Casimir effect, was considered by
Muñoz-Castañeda and Guilarte \cite{Castaneda-Guilarte-2015}, resulting
in a generalization of (\ref{eq:Barton}), done by adding a 
$\delta^{\prime}$ term in the potential, namely 
\begin{equation}
\mathcal{L}=(1/2)[(\partial_{t}\phi)^{2}-(\partial_{x}\phi)^{2}]-[\mu\delta(x)+\lambda\delta^{\prime}(x)]\phi^{2}(t,x),
\label{eq:Munoz}
\end{equation}
where $\lambda$ is dimensionless. Following Refs. \cite{Castaneda-Guilarte-2015,Kurasov-1996,Gadella-2009}, we can find that the transmission and reflection coefficients are given by (see the Appendix for details)
\begin{equation}
s_{\pm}(\omega)=\frac{\omega(1-\lambda^{2})}{\omega(\lambda^{2}+1)+i\mu},\enskip r_{\pm}(\omega)=\frac{\pm2\omega\lambda-i\mu}{\omega(\lambda^{2}+1)+i\mu}.
\label{eq:s-and-r}
\end{equation}
Note that, differently of (\ref{eq:Barton}), in this case $r_{+}(\omega)\neq r_{-}(\omega)$.
Moreover, $\lambda\rightarrow-\lambda$
is equivalent to change the mirror properties from left to right: $r_{\pm}(\omega)\rightarrow r_{\mp}(\omega)$.
For $\lambda = 1$ the mirror is perfectly reflecting
[$s_{\pm}(\omega)\rightarrow0$] and the following boundary conditions are
imposed to the field:
\begin{eqnarray}
\phi_{+}(t,0^{+})-(2/\mu)\partial_{x}\phi_{+}(t,0^{+}) & = & 0,\label{eq:RobinBC}\\
\phi_{-}(t,0^{-}) & = & 0.\label{eq:Dirichlet-BC}
\end{eqnarray}
These are, respectively, the Robin and the Dirichlet boundary conditions.

In the present paper, we investigate the DCE for a real massless scalar field in $1+1$ dimensions in the presence of a $\delta-\delta^{\prime}$ moving mirror, computing the spectrum and the total rate of created particles. 
The influence of the coupling constants $\mu$ and $\lambda$ to the particle production is described and, in the limit of a perfect mirror, the results are compared with those for the particle creation with Robin conditions found in the literature \cite{Mintz-Farina-MaiaNeto-Rodrigues-2006-I}.

This paper is organized as follows. In Sec. \ref{sec:Scattering-approach},
we use the scattering approach \cite{Jaekel-Reynaud-1992,Lambrecht-Jaekel-Reynaud-1996}
to outline  general aspects of the spectrum of created particles for a partially reflecting
mirror, with arbitrary scattering coefficients, considering a typical
function for the movement of the mirror. In Sec. \ref{sec:Particle-creation-phenomenon},
we consider specifically the $\delta-\delta^{\prime}$ mirror and
compute the spectrum and the total rate of created particles. 
The final remarks are presented in Sec. \ref{sec:Final-Remarks}.

%%%%%%%%%%%%%%%%%%%%%%%%%%%%%%%%%%%%%%%%%%%%%%%%%%%%%%%%%%%%%%%%%%%%%%%%%%%%%%%%%%%%%%%%%%%%%%%%%%%%%%%
\section{General framework of the scattering approach\label{sec:Scattering-approach}}

Let us start by considering a generic mirror at rest, for simplicity,
at $x=0$. The field is then written as 
\begin{equation}
\phi(t,x)=\Theta(x)\phi_{+}(t,x)+\Theta(-x)\phi_{-}(t,x),
\label{phi-00}
\end{equation}
where $\Theta(x)$ is the Heaviside step function. Also, $\phi_{+}$
and $\phi_{-}$ obey the massless Klein-Gordon equation, $(\partial_{x}^{2}-\partial_{t}^{2})\phi_{\pm}(t,x)=0$.
Thus they are the sum of two freely counterpropagating fields, 
\begin{equation}
\phi_{+}(t,x)=\int\frac{\mathrm{d}\omega}{\sqrt{2\pi}}\left[\varphi_{\text{out}}(\omega)\text{e}^{i\omega x}+\psi_{\text{in}}(\omega)\text{e}^{-i\omega x}\right]\text{e}^{-i\omega t},\label{eq:A08}
\end{equation}
\begin{equation}
\phi_{-}(t,x)=\int\frac{\mathrm{d}\omega}{\sqrt{2\pi}}\left[\varphi_{\text{in}}(\omega)\text{e}^{i\omega x}+\psi_{\text{out}}(\omega)\text{e}^{-i\omega x}\right]\text{e}^{-i\omega t},\label{eq:A09}
\end{equation}
where the labels ``in'' and ``out'' indicate the amplitudes of
the incoming and outgoing fields respectively.

The presence of the mirror does not affect the incoming fields, thus
it is straightforward to show that 
\begin{equation}
\varphi_{\text{in}}(\omega)=(2\left|\omega\right|)^{-1/2}\left[\Theta(\omega)a_{L}(\omega)+\Theta(-\omega)a_{L}^{\dagger}(-\omega)\right],\label{eq:A04}
\end{equation}
\begin{equation}
\psi_{\text{in}}(\omega)=(2\left|\omega\right|)^{-1/2}\left[\Theta(\omega)a_{R}(\omega)+\Theta(-\omega)a_{R}^{\dagger}(-\omega)\right],\label{eq:A05}
\end{equation}
where $a_j(\omega)$ and $a_j^{\dagger}(\omega)$ ($j=L,R$) are annihilation and creation operators, obeying the relation
$[a_i(\omega),a_j^{\dagger}(\omega^{\prime})]=\delta(\omega-\omega^{\prime})\delta_{ij}$.
The outgoing fields correspond to the incoming ones scattered by the
mirror. They can be linearly obtained by \cite{Jaekel-Reynaud-1991,Jaekel-Reynaud-1992}
\begin{equation}
\Phi_{\text{out}}(\omega)=S(\omega)\Phi_{\text{in}}(\omega),\label{eq:13-1}
\end{equation}
where 
\begin{equation}
\Phi_{\text{out}}(\omega)=\left(\begin{array}{c}
\varphi_{\text{out}}(\omega)\\
\psi_{\text{out}}(\omega)
\end{array}\right),\enskip\Phi_{\text{in}}(\omega)=\left(\begin{array}{c}
\varphi_{\text{in}}(\omega)\\
\psi_{\text{in}}(\omega)
\end{array}\right),\label{eq:A10}
\end{equation}
and $S(\omega)$ is a $2\times2$ matrix denominated scattering matrix
($S$-matrix). 

In the particular case of a perfectly reflecting mirror,
the outgoing fields correspond just to the reflected incoming ones, multiplied by
a phase term (which depends on the boundary condition imposed by the
mirror), namely
\begin{equation}
\varphi_{\text{out}}(\omega)=\text{e}^{i\theta_{+}(\omega)}\psi_{\text{in}}(\omega),\quad\psi_{\text{out}}(\omega)=\text{e}^{i\theta_{-}(\omega)}\varphi_{\text{in}}(\omega).\label{eq:A51-1}
\end{equation}
Thus, for a perfect mirror,
\begin{equation}
S(\omega)=\left(\begin{array}{cc}
0 & \mathrm{e}^{i\theta_{+}(\omega)}\\
\mathrm{e}^{i\theta_{-}(\omega)} & 0
\end{array}\right),\label{eq:perfect-scattering-matrix}
\end{equation}
with $\theta_{+}$ and $\theta_{-}$ being the phases. 

In the general case of a partially reflecting mirror, the $S$-matrix is generalized
to 
\begin{equation}
S(\omega)=\left(\begin{array}{cc}
s_{+}(\omega) & r_{+}(\omega)\\
r_{-}(\omega) & s_{-}(\omega)
\end{array}\right),\label{eq:matriz-espalhamento}
\end{equation}
where $r_{\pm}(\omega)$ and $s_{\pm}(\omega)$ are the reflection and
transmission coefficients, which are assumed to obey the following
conditions \cite{Jaekel-Reynaud-1991}. Since the field is real, the
elements of $S(\omega)$ are also real in the temporal domain, therefore
$S(-\omega)=S^{*}(\omega)$. As a consequence of the commutation
rule $[\phi(t,x),\phi(t,y)]=0$, the $S$-matrix is unitary, namely
$S(\omega)S^{\dagger}(\omega)=\mathbb{I}$, which means that there
is not dissipative effects in the mirror (for lossy mirrors
the $S$-matrix is not unitary and the quantization is changed
\cite{MIT}). As a consequence
of the commutation rule $[\phi(t,x),\dot{\phi}(t,y)]=i\delta(x-y)$
the $S$-matrix is causal, which means that $s_{\pm}(\omega)$ and
$r_{\pm}(\omega)$ vanishes in the temporal domain for $t<0$ \cite{Jaekel-Reynaud-1991,Moysez}.
This causality condition is fulfilled when $S(\omega)$ is analytic
for $\mathrm{Im}(\omega)>0$. 
The coefficients for $\delta-\delta^{\prime}$ mirrors [Eq. (\ref{eq:s-and-r})] satisfy all these properties.

Now, we shall consider the scattering for a moving mirror. The position
of the mirror is represented by $x=q(t)$, and the movement is set
nonrelativistic, $\left|\dot{q}(t)\right|\ll1$, and limited by a
small value $\epsilon$, $q(t)=\epsilon g(t)$ with $|g(t)|\le1$.
We consider inertial frames where the mirror is instantaneously at
rest (tangential frames) and the scattering is assumed to be \cite{Jaekel-Reynaud-1992}
\begin{equation}
\Phi_{\text{out}}^{\prime}(\omega)=S(\omega)\Phi_{\text{in}}^{\prime}(\omega),\label{eq:Q15}
\end{equation}
where the prime superscript means that this relation is taken in the
tangential frame. 
In order to find $\Phi_{\text{out}}^{\prime}$ and $\Phi_{\text{in}}^{\prime}$
in the laboratory frame, we start from the relation $\tilde{\Phi}^{\prime}(t^{\prime},0)=\tilde{\Phi}(t,\epsilon g(t))$,
or 
\begin{equation}
\tilde{\Phi}^{\prime}(t^{\prime},0)=\left[1-\epsilon g(t)\eta\partial_{t}\right]\tilde{\Phi}(t,0)+\mathcal{O}(\epsilon^{2}),
\end{equation}
where 
\begin{equation}
\tilde{\Phi}(t,x)=\left(\begin{array}{c}
\tilde{\varphi}(t-x)\\
\tilde{\psi}(t+x)
\end{array}\right),
\end{equation}
$\tilde{\varphi}$ and $\tilde{\psi}$ are the components of the field
in the temporal domain, and $\eta=\mathrm{diag}(1,-1)$. Moreover,
$\mathrm{d}t^{\prime}=\mathrm{d}t+\mathcal{O}(\epsilon^{2})$. Therefore,
neglecting the terms $\mathcal{O}(\epsilon^{2})$, $t^{\prime}$ can
be replaced by $t$, namely $\tilde{\Phi}^{\prime}(t,0)=\left[1-\epsilon g(t)\eta\partial_{t}\right]\tilde{\Phi}(t,0)$ which,
in the Fourier domain, reads 
\begin{equation}
\Phi^{\prime}(\omega)=\Phi(\omega)+i\epsilon\eta\int\frac{\mathrm{d}\Omega}{2\pi}\Omega G(\omega-\Omega)\Phi(\Omega),\label{eq:23-1-1}
\end{equation}
where $G(\omega)$ is the Fourier transform of $g(t)$, and $\Phi(\omega)$
and $\Phi^{\prime}(\omega)$ are short notations for $\Phi(\omega,0)$
and $\Phi^{\prime}(\omega,0)$. The application of Eq. (\ref{eq:23-1-1}),
duly labeled with \emph{out} and \emph{in}, in Eq. (\ref{eq:Q15})
leads to 
\begin{eqnarray}
\Phi_{\text{out}}(\omega) & = & S(\omega)\Phi_{\text{in}}(\omega)+\int\frac{\mathrm{d}\Omega}{2\pi}\mathcal{S}(\omega,\Omega)\Phi_{\text{in}}(\Omega),\label{eq:26-1}\\
\nonumber \\
\mathcal{S}(\omega,\Omega) & = & i\epsilon\Omega G(\omega-\Omega)\left[S(\omega)\eta-\eta S(\Omega)\right].\label{eq:24}
\end{eqnarray}
Therefore, the movement of the mirror led to a first-order correction
to the $S$-matrix. The relation (\ref{eq:26-1}) enables us to compute
the spectrum of created particles in the following.

The total number of created particles for the problem under investigation
is 
\begin{equation}
\mathcal{N}=\int_{0}^{\infty}\mathrm{d\omega}\,N(\omega),\label{eq:total-number}
\end{equation}
where $N(\omega)$ is the spectral distribution of created particles,
given by \cite{Jaekel-Reynaud-1992,Lambrecht-Jaekel-Reynaud-1996}
\begin{equation}
N(\omega)=2\omega\,\mathrm{Tr}\left[\left\langle 0_{\text{in}}\left|\Phi_{\text{out}}(-\omega)\Phi_{\text{out}}^{\mathrm{T}}(\omega)\right|0_{\text{in}}\right\rangle \right],\label{eq-lambrecht-prl-1996}
\end{equation}
and the incoming fields are assumed to be in the vacuum state. Inserting
Eq. (\ref{eq:26-1}) into Eq. (\ref{eq-lambrecht-prl-1996}) and considering
the formula 
\begin{equation}
\left\langle 0_{\text{in}}\left|\Phi_{\text{in}}(\omega)\Phi_{\text{in}}^{\mathrm{T}}(\omega^{\prime})\right|0_{\text{in}}\right\rangle =(\pi/\omega)\delta(\omega+\omega^{\prime})\Theta(\omega),
\end{equation}
obtained from Eqs. (\ref{eq:A04}) and (\ref{eq:A05}), it is straightforward
to show that 
\begin{equation}
N(\omega)=\frac{1}{2\pi}\int_{0}^{\infty}\frac{\mathrm{d}\Omega}{2\pi}\frac{\omega}{\Omega}\mathrm{Tr}\left[\mathcal{S}\left(\omega,-\Omega\right)\mathcal{S}^{\dagger}(\omega,-\Omega)\right].\label{eq:42}
\end{equation}
Substituting Eq. (\ref{eq:24}) in (\ref{eq:42}), we get 
\begin{equation}
N(\omega)=\frac{4\epsilon^{2}}{\pi}\int_{0}^{\infty}\frac{\mathrm{d}\Omega}{2\pi}\omega\Omega\left|G(\omega+\Omega)\right|^{2}\Lambda(\omega,\Omega),\label{eq:67}
\end{equation}
\begin{eqnarray}
\Lambda(\omega,\Omega) & = & \frac{1}{4}\Re\big[1+r_{+}(\omega)r_{+}(\Omega)-s_{+}(\omega)s_{+}(\Omega)+\nonumber \\
 &  & \qquad1+r_{-}(\omega)r_{-}(\Omega)-s_{-}(\omega)s_{-}(\Omega)\big],\label{eq:lambda}
\end{eqnarray}
where $0\le\Lambda(\omega,\Omega)\le1$.
Equation (\ref{eq:67}) gives us the spectrum of created particles
if the scattering coefficients and the motion function of the mirror
are provided.

Henceforth, we shall consider the following typical motion for the
mirror 
\begin{equation}
g(t)=\cos(\omega_{0}t)\exp(-\left|t\right|/\tau),\label{eq:movement}
\end{equation}
where $\tau$ is the time for which the oscillations occur effectively,
and $\omega_{0}$ is the characteristic frequency of oscillation.
In addition, we shall consider $\omega_{0}\tau\gg1$ (monochromatic
limit \cite{Silva-Braga-Rego-Alves-2015}), what leads to an effective spatially symmetric movement.
The Fourier transform of $g(t)$ is 
\begin{equation}
G(\omega)=\frac{2\tau\left[1+\tau^{2}\left(\omega^{2}+\omega_{0}^{2}\right)\right]}{\left[1+(\omega-\omega_{0})^{2}\tau^{2}\right]\left[1+(\omega+\omega_{0})^{2}\tau^{2}\right]}.\label{eq:41-1}
\end{equation}
It presents sharp peaks around $\omega=\pm\omega_{0}$, so that in
the monochromatic limit \cite{Silva-Braga-Rego-Alves-2015}
\begin{equation}
\lim_{\tau\rightarrow\infty}\left|G(\omega)\right|^{2}/\tau=(\pi/2)\left[\delta(\omega-\omega_{0})+\delta(\omega+\omega_{0})\right].\label{eq:C85}
\end{equation}
Using the Eq. (\ref{eq:C85}), we analyze the behavior of $N(\omega)/\tau$
in the monochromatic limit.

Substituting Eq. (\ref{eq:C85}) in (\ref{eq:67}) we obtain 
\begin{equation}
N(\omega)/\tau=(\epsilon^{2}/\pi)\omega(\omega_{0}-\omega)\Lambda(\omega,\omega_{0}-\omega)\Theta(\omega_{0}-\omega).
\label{eq:C84}
\end{equation}
Notice that, independently of the scattering coefficients, there are
not created particles with frequency $\omega>\omega_{0}$. Moreover,
the spectrum is symmetrical with respect to $\omega=\omega_{0}/2$,
since it is invariant under the change $\omega\rightarrow\omega_{0}-\omega$.
This is interpreted as a signature of the fact that particles are
created in pairs: for each particle created with a frequency $\omega$
there is another with frequency $\omega_{0}-\omega$ \cite{Mintz-Farina-MaiaNeto-Rodrigues-2006-I,Silva-Braga-Rego-Alves-2015,Rego-Mintz-Farina-Alves-PRD-2013,Silva-Farina-2011}.

The scattering for a perfect mirror is described by Eq. (\ref{eq:perfect-scattering-matrix})
and, for this case, Eq. (\ref{eq:lambda}) becomes 
\begin{equation}
\Lambda(\omega,\Omega)=\frac{1}{4}\Re\left[2+\mathrm{e}^{i\theta_{+}(\omega)}\mathrm{e}^{i\theta_{+}(\Omega)}+\mathrm{e}^{i\theta_{-}(\omega)}\mathrm{e}^{i\theta_{-}(\Omega)}\right].\label{eq:P87}
\end{equation}
Particularly, for the Neumann and Dirichlet boundary conditions, corresponding
respectively to $\theta_{\pm}(\omega)=0$ and $\theta_{\pm}(\omega)=\pi$,
we get $\Lambda(\omega,\Omega)=1$. Therefore, the spectra for these
cases are not only identical, but they also correspond to the cases
where the greatest number of particles is produced. On the other hand,
for $\theta_{\pm}(\omega)=\pi/2$, it follows that $\Lambda(\omega,\Omega)=0$
and, consequently, no particles would be created. Thus, one can say
that the phase $\pi/2$ results, in the context of the particle creation,
in a complete decoupling between the field and the mirror. When
the mirror imposes the Robin boundary condition (\ref{eq:RobinBC})
to the field, it is straightforward to show that $\theta_{+}(\omega)=2\arctan(2\omega/\mu)$
and, for the particular value $2\omega_{0}/\mu\approx2.2$,  it occurs
a very strong inhibition of the particle production in $1+1$ \cite{Mintz-Farina-MaiaNeto-Rodrigues-2006-I,Silva-Braga-Rego-Alves-2015}
and in $3+1$ \cite{Rego-Mintz-Farina-Alves-PRD-2013} dimensions.
%
%%%%%%%%%%%%%%%%%%%%%%%%%%%%%%%%%%%%%

\section{Particle creation phenomenon for a $\delta-\delta^{\prime}$ mirror}
\label{sec:Particle-creation-phenomenon}
\begin{figure}[t]
\begin{centering}
\includegraphics[width=0.95\columnwidth]{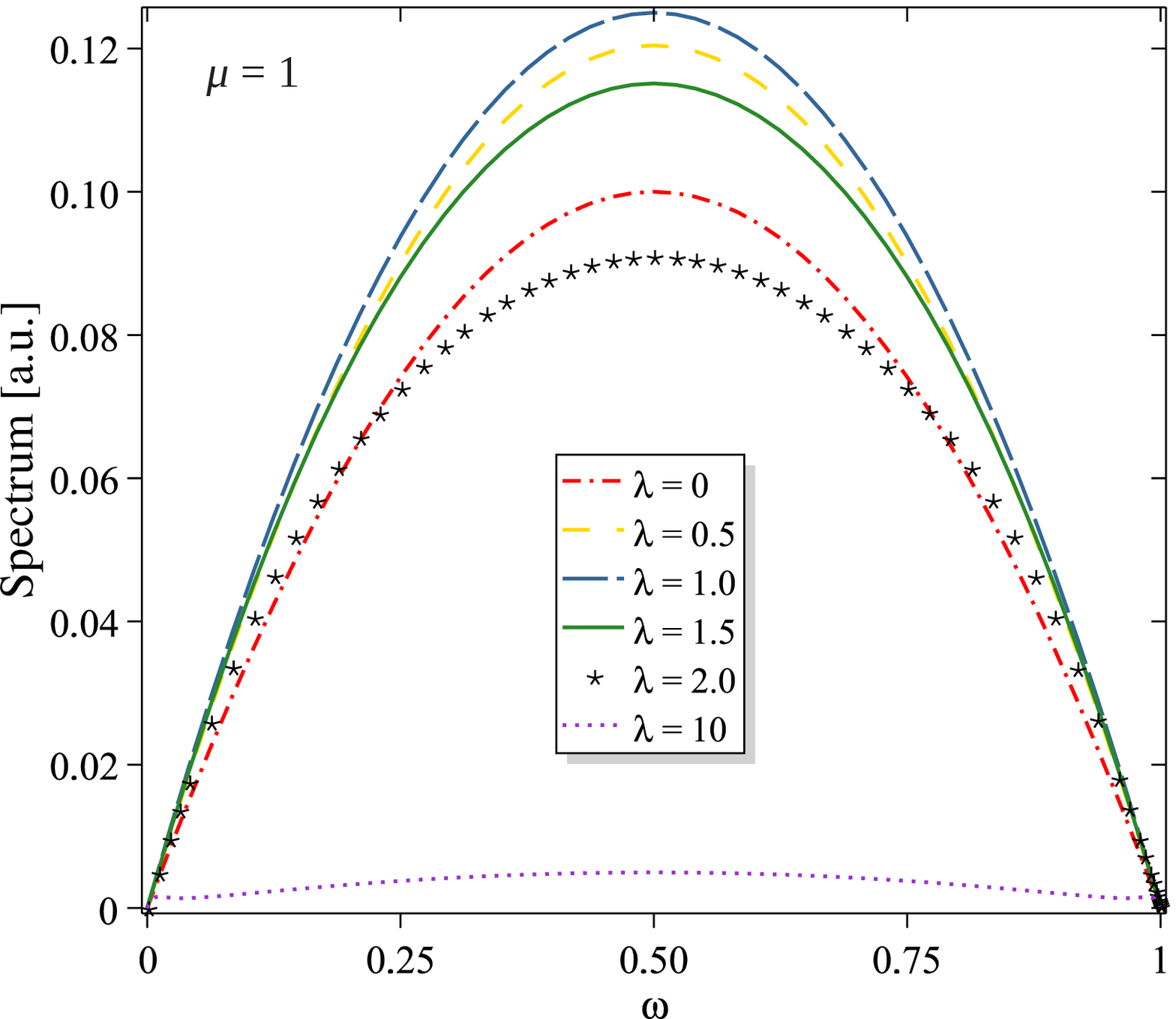} 
\end{centering}
\caption{$(\epsilon^{2}\tau/\pi)^{-1}\times N_{-}(\omega)$ as a function of
$\omega$, with $\mu=\omega_0=1$ and several values for $\lambda$.}
\label{spectrum-lambda} 
\end{figure}
\begin{figure}[t]
\begin{centering}
\includegraphics[width=0.95\columnwidth]{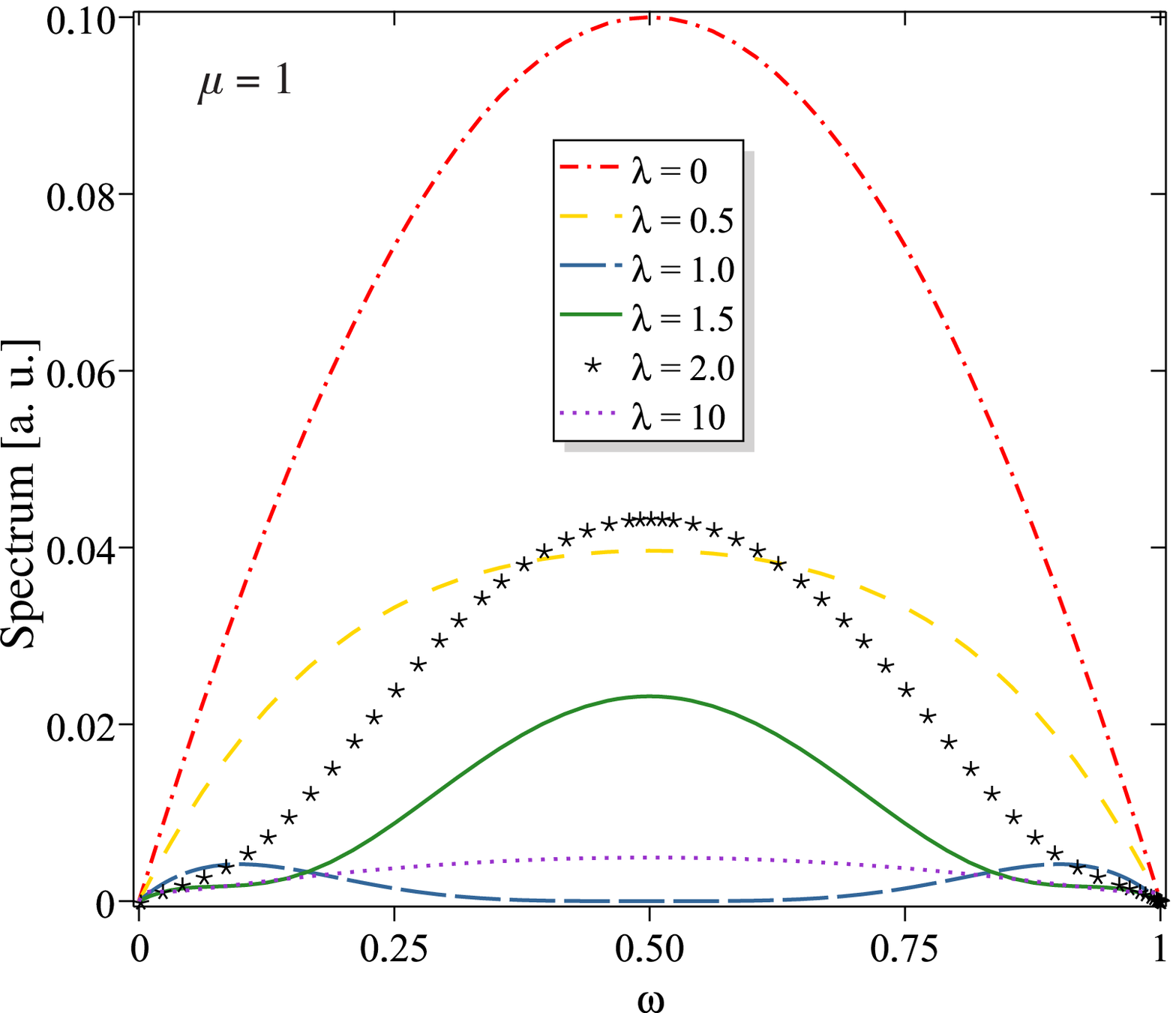}
\end{centering}
\caption{$(\epsilon^{2}\tau/\pi)^{-1}\times N_{+}(\omega)$ as a function of
$\omega$, with $\mu=\omega_0=1$ and several values for $\lambda$.}
\label{spectrum-lambda-1} 
\end{figure}
Before calculating the DCE for the $\delta-\delta^{\prime}$ model, we briefly discuss some general aspects of this model.
The Dirichlet condition (\ref{limite-dirichlet-barton}) is related to perfect mirrors [$s_{\pm}(\omega)=0$]
with frequency-independent reflection coefficients $r_{\pm}(\omega)=-1$.
On the other hand, real mirrors are naturally transparent at high frequencies \cite{Moore-1970,Casimir-1948}.
A way to model partially reflecting mirrors is via Dirac $\delta$ potentials (\ref{eq:Barton}),
what gives the transmission and reflection coefficients shown in Eq. (\ref{eq:s-and-r-barton}) which obey 
the condition of transparency at high frequencies: $\lim_{\omega\rightarrow\infty}s_{\pm}(\omega)=1$. 
The model (\ref{eq:Barton}), in the limit of a perfect mirror ($\mu\rightarrow\infty$), gives the Dirichlet boundary condition in both sides of the mirror (\ref{limite-dirichlet-barton}).
The $\delta-\delta^{\prime}$ model (\ref{eq:Munoz})
simulates partially reflecting mirrors with
transmission and reflection
coefficients given by Eq. (\ref{eq:s-and-r}). This model has the interesting property
that in the limit of a perfect mirror ($\lambda=1$) it gives 
the Dirichlet (\ref{eq:Dirichlet-BC}) and Robin (\ref{eq:RobinBC}) boundary conditions, being the Neumann condition a particular case 
obtained considering $\mu=0$ in (\ref{eq:RobinBC}). Although the model (\ref{eq:Munoz}) with $\lambda\neq 1$ represents a partially reflecting mirror and, in this way, is more realistic (in comparison with the case $\lambda=1$), it is not transparent at high frequencies, so that
\begin{equation}
\lim_{\omega\rightarrow\infty}s_{\mathrm{\pm}}(\omega)=\frac{1-\lambda^{2}}{1+\lambda^{2}}<1,\quad(\lambda\neq0).
\label{eq:hhh}
\end{equation}
Moreover, for $\mu=0$ the scattering coefficients for the $\delta-\delta^{\prime}$ mirror (\ref{eq:s-and-r}) are independent of the frequency,
\begin{equation}
s_{\pm}(\omega)=\frac{1-\lambda^{2}}{1+\lambda^{2}},\quad r_{\pm}(\omega)=\frac{\pm2\lambda}{1+\lambda^{2}}
\end{equation}
(a similar model with frequency-independent scattering coefficients is found in Ref. \cite{Lambrecht-Jaekel-Reynaud-1998}).
However, this behavior at high frequencies ($\omega\gg\omega_0$) does not affect the application of the formula (\ref{eq:C84}) for the $\delta-\delta^{\prime}$ model (\ref{eq:Munoz}), since the DCE, in the approximation assumed in the present paper, just depends on the scattering coefficients for $\omega<\omega_0$ [notice the Heaviside function in Eq. (\ref{eq:C84})].

We can write $\Lambda$ [Eq. (\ref{eq:lambda})] as $\Lambda=\Lambda_{+}+\Lambda_{-}$, where 
\begin{equation}
\Lambda_{\pm}(\omega,\Omega)=\frac{1}{4}\Re\left[1+r_{\pm}(\omega)r_{\pm}(\Omega)-s_{\pm}(\omega)s_{\pm}(\Omega)\right].
\label{eq:Lambda-m-m}
\end{equation}
We can also write $N(\omega)=N_{+}(\omega)+N_{-}(\omega)$, with 
\begin{equation}
N_{\pm}(\omega)=N_{\mathrm{D}}(\omega)\times\Lambda_{\pm}(\omega,\omega_{0}-\omega),\label{eq:spectrum-m-m}
\end{equation}
where $N_{\mathrm{D}}(\omega)/\tau=(\epsilon^{2}/\pi)\omega(\omega_{0}-\omega)\Theta(\omega_{0}-\omega)$
is the spectrum for the Dirichlet (or Neumann) case,
and $N_{+}$ ($N_{-}$) is the spectrum in the right (left) side of
the mirror. In the same way, $\mathcal{N}=\mathcal{N}_{+}+\mathcal{N}_{-}$, where
$\mathcal{N}_{+}$ ($\mathcal{N}_{-}$) is the total number of created particles in the right (left) side of the mirror.

Substituting the scattering coefficients of the $\delta-\delta^{\prime}$
mirror given by Eq. (\ref{eq:s-and-r}) in Eq. (\ref{eq:Lambda-m-m}),
we obtain
\begin{equation}
\Lambda_{\pm}=\Re\left[\frac{2\xi\alpha^{2}(1-\xi)\lambda^{2}-1/2+i\alpha(\lambda\mp1)^{2}/4}{\xi\alpha^{2}(1-\xi)(\lambda^{2}+1)^{2}-1+i\alpha(\lambda^{2}+1)}\right],\label{eq:lambda-m-m}
\end{equation}
where we have defined the dimensionless variables 
\begin{equation}
\xi=\omega/\omega_{0}\quad\text{and}\quad\alpha=\omega_{0}/\mu.
\end{equation}
From Eqs. (\ref{eq:spectrum-m-m}) and (\ref{eq:lambda-m-m}) we see that the spectra for each side of the mirror are different,
what is a consequence of the fact that the scattering on each side are not the same.
The change $\lambda\rightarrow-\lambda$ is equivalent to $\Lambda_{+}\rightarrow\Lambda_{-}$
(or $N_{+}\rightarrow N_{-}$).

The case $\lambda=1$ corresponds to the spectrum of a perfectly reflecting 
$\delta-\delta^{\prime}$ mirror, where 
\begin{eqnarray}
\Lambda_{+} & = & \frac{1}{2}\frac{\left[1-4\alpha^{2}(1-\xi)\xi\right]^{2}}{\left(1+4\alpha^{2}\xi^{2}\right)\left[1+4\alpha^{2}(1-\xi)^{2}\right]},\nonumber \\
 \Lambda_{-}& = & \frac{1}{2},
\label{eq:lambda-1}
\end{eqnarray}
with $\Lambda_{-}$ corresponding to the parabolic spectrum (just one side) for
a Dirichlet mirror (long-dashed line in Fig. \ref{spectrum-lambda}),
in agreement with Ref. \cite{Lambrecht-Jaekel-Reynaud-1996},
whereas $\Lambda_{+}$ corresponds to the spectrum for a mirror imposing
the Robin boundary condition (long-dashed line in Fig. \ref{spectrum-lambda-1}), in agreement with Ref.
\cite{Mintz-Farina-MaiaNeto-Rodrigues-2006-I}. For $\Lambda_{+}$,
when $\mu=\omega_{0}$ it follows that $\theta_{+}(\omega_{0}/2)=\pi/2$ and,
from Eqs. (\ref{eq:C84}) and (\ref{eq:P87}),  
$N_{+}(\omega_{0}/2)=0$, 
which results in a strong inhibition of the particle production (as discussed in Ref. \cite{Mintz-Farina-MaiaNeto-Rodrigues-2006-I}).
When $\mu=0$ it follows that $\theta_{+}(\omega)= 0$, corresponding to the spectrum 
for a Dirichlet mirror, whereas for $\mu\rightarrow\infty$ the phase becomes $\theta_{+}(\omega)= \pi$, resulting in the spectrum 
produced by a Neumann mirror.

For the case $\lambda=0$,
\begin{equation}
\Lambda_{+}=\Lambda_{-}=\frac{1+\alpha^{2}\left[1-2\xi\left(1-\xi\right)\right]/2}{(1+\alpha^{2}\xi^{2})\left[1+\alpha^{2}(1-\xi)^{2}\right]},
\label{pure-delta-lambda-m-m}
\end{equation}
what corresponds to a pure $\delta$ mirror, 
which produces identical spectra for both sides,
increasing monotonically with $\mu$ and going asymptotically to the Dirichlet
spectrum when $\mu\rightarrow\infty$.

In Figs. \ref{spectrum-lambda}
and \ref{spectrum-lambda-1} we compare the behaviors of $\mathcal{N}_{-}$ and $\mathcal{N}_{+}$
(the areas under the curves), for $\mu=1$.
From $\lambda=0$ up to $\lambda=1$,
we see in Fig. \ref{spectrum-lambda} an increase
of $\mathcal{N}_{-}$, whereas in Fig. \ref{spectrum-lambda-1} we see a 
decrease of $\mathcal{N}_{+}$.
When $\lambda=1$, we see that $\mathcal{N}_{-}$ is much greater than 
$\mathcal{N}_{+}$.
From $\lambda=1$ to $\lambda=2$, we see in Fig. \ref{spectrum-lambda} a decrease
of $\mathcal{N}_{-}$ and, in Fig. \ref{spectrum-lambda-1}, the opposite
behavior for $\mathcal{N}_{+}$.
From $\lambda=2$ to $\lambda=10$, both
$\mathcal{N}_{-}$ and $\mathcal{N}_{+}$ diminish, in according to Eq. (\ref{eq:lambda-m-m}),
from which we can conclude that $\lim_{\lambda\rightarrow\infty}\mathcal{N}_{\pm}=0$. 

Next, we turn to investigate the total number of created particles
$\mathcal{N}$. 
Substituting Eq. (\ref{eq:lambda-m-m})
in (\ref{eq:spectrum-m-m}) and then in (\ref{eq:total-number}),
we obtain
\begin{widetext}
\begin{equation}
\mathcal{N}/\tau=\frac{\epsilon^{2}\omega_{0}^{3}}{6\pi}\times\frac{\mathcal{A}(\alpha,\lambda)+\mathcal{B}(\alpha,\lambda)\ln\left[\alpha^{2}(1+\lambda^{2})^{2}+1\right]+\mathcal{C}(\alpha,\lambda)\arctan\left[\alpha(1+\lambda^{2})\right]}{\alpha^{3}(1+\lambda^{2})^{5}\left[\alpha^{2}(1+\lambda^{2})^{2}+4\right]},\label{eq:total-number-1}
\end{equation}
\begin{eqnarray}
\mathcal{A}(\alpha,\lambda) & = & 4\alpha^{5}\lambda^{2}(1+\lambda^{2})^{5}-24\alpha(\lambda^{2}-1)^{2}(1+\lambda^{2})-6\alpha^{3}(1+\lambda^{2})^{5}+40\alpha^{3}\lambda^{2}(\lambda^{2}+1)^{3},\\
\nonumber \\
\mathcal{B}(\alpha,\lambda) & = & 3\alpha^{3}\left[(\lambda^{2}-1)^{2}-4\lambda^{2}\right](\lambda^{2}+1)^{3}+12\alpha\left[(\lambda^{2}-1)^{2}-2\lambda^{2}\right](\lambda^{2}+1),\\
\nonumber \\
\mathcal{C}(\alpha,\lambda) & = & 6\alpha^{2}(\lambda^{2}+1)^{4}+24(\lambda^{2}-1)^{2}.
\end{eqnarray}

\end{widetext}

\noindent When $\mu\rightarrow\infty$, the total rate of created particles
for a Dirichlet mirror is recovered,
namely $\mathcal{N}_{\mathrm{D}}/\tau=\epsilon^{2}\omega_{0}^{3}/(6\pi)$
(in agreement with Ref. \cite{Lambrecht-Jaekel-Reynaud-1996}).
From Eq. (\ref{eq:total-number-1}), we see that $\mathcal{N}/\mathcal{N}_{\mathrm{D}}\leq 1$ or,
in other words, the Dirichlet case is a situation of a maximum number of created particles.

\begin{figure}[t]
\begin{centering}
\includegraphics[width=0.99\columnwidth]{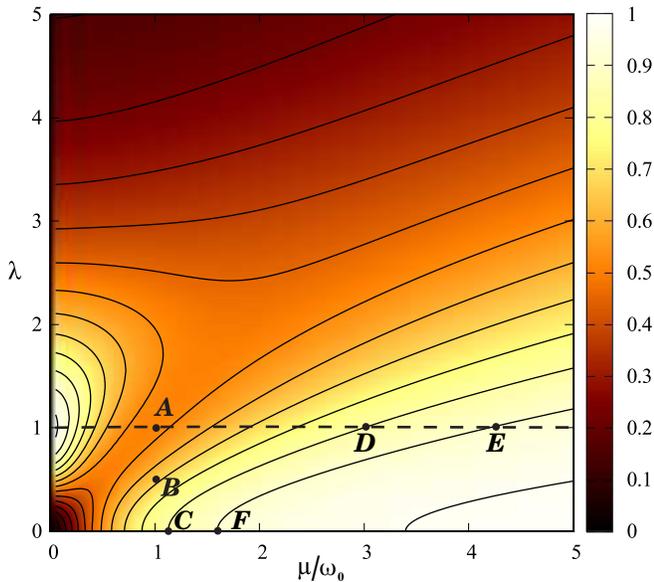}
\end{centering}
\caption{$\mathcal{N}/\mathcal{N}_{\mathrm{D}}$ as a function
of $\mu/\omega_0$ and $\lambda$. The dashed line ($\lambda=1$) represents the region where
the mirror is perfectly reflecting. The points $A$, $D$ and $E$ illustrate perfect $\delta-\delta^{\prime}$ mirrors. 
The points $C$ and $F$ indicate partially transparent $\delta$ mirrors.
The point $B$ shows a partially transparent $\delta-\delta^{\prime}$ mirror. The solid lines are the level curves.
}
\label{fig:Total-number} 
\end{figure}

Results using (\ref{eq:total-number-1}) are shown in Fig. \ref{fig:Total-number}. 
For $\lambda=0$ (horizontal $\mu$-axis), what corresponds
to a pure $\delta$ mirror, the enhancement of the transparency (by reducing $\mu$)
leads to a monotonic reduction of $\mathcal{N}/\mathcal{N}_{\mathrm{D}}$, being
$\lim_{\mu\rightarrow 0}\mathcal{N}/\mathcal{N}_{\mathrm{D}}=0$. 
For $\lambda=1$ (dashed line in Fig. \ref{fig:Total-number}), the mirror is perfectly
reflecting, and the field satisfies the Robin (\ref{eq:RobinBC})
and Dirichlet (\ref{eq:Dirichlet-BC}) boundary conditions, each one on
a given side of the mirror, being the total rate
not monotonic with $\mu/\omega_0$. 
The point $\mu=0$ and $\lambda=1$ in Fig. \ref{fig:Total-number} corresponds
to the case of the Neumann boundary condition ($\mathcal{N}/\mathcal{N}_{\mathrm{D}}=1$),
being Dirichlet and Neumann the cases of maximum particle creation rate.
Finally, $\lim_{\mu\rightarrow \infty}\mathcal{N}/\mathcal{N}_{\mathrm{D}}=1$
(not depending on the value of $\lambda$).

\begin{figure}[t]
\begin{centering}
\includegraphics[width=0.96\columnwidth]{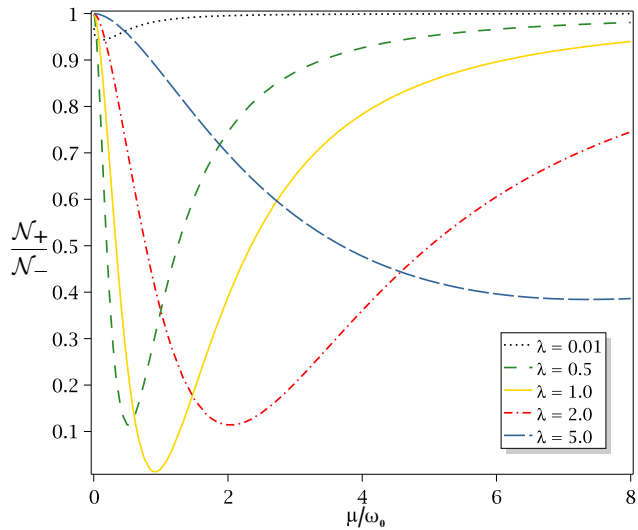}
\end{centering}
\caption{$\mathcal{N}_{+}/\mathcal{N}_{-}$ as a function
of $\mu/\omega_0$, for several values of $\lambda$.}
\label{fig:Ratio} 
\end{figure}

For a pure $\delta$ mirror ($\lambda = 0$), the reflectivity $|r_{\pm}(\omega)|$
and the phase $\arg [r_{\pm}(\omega)]$ are constrained so that the rate of particles always increases with the 
enhancement of the reflectivity and, therefore, the greatest number of particles is obtained for a perfect mirror ($\mu \rightarrow \infty$).
In the $\delta-\delta^{\prime}$ case, the constraint between $|r_{\pm}(\omega)|$
and $\arg [r_{\pm}(\omega)]$ enables transparent mirrors ($\lambda\neq 1$ and $\mu<\infty$) creating more particles
than a perfect one ($\lambda=1$).
For example, let us consider the points $A$ $(\mu/\omega_0=1,\lambda=1)$ and $B$ $(\mu/\omega_0=1,\lambda=1/2)$ in Fig. \ref{fig:Total-number}. The point
$A$ represents a perfect mirror {[}$|r_{\pm}(\omega)|=1${]},
whereas the point $B$ represents a partially reflecting one {[}$|r_{\pm}(\omega)|<1${]}.
As shown in Fig. \ref{fig:Total-number}, the change $A\rightarrow B$ enhances the transparency, but increases the number of produced particles. 
This can be also visualized with the help of the Figs. \ref{spectrum-lambda} and \ref{spectrum-lambda-1}.
In Fig. \ref{spectrum-lambda}, the change $A\rightarrow B$ 
[shown by the transition from the long-dashed line ($\lambda=1$) to the space-dashed one ($\lambda=1/2$)]
corresponds to a variation $\Delta {\mathcal N}_{-}<0$ in the particle production (difference between the areas under 
the long-dashed and  space-dashed curves), whereas
in Fig. \ref{spectrum-lambda-1}, $A\rightarrow B$
corresponds to a variation $\Delta  {\mathcal N}_{+}>0$.
The total variation is $\Delta  {\mathcal N}_{+}+\Delta  {\mathcal N}_{-}>0$, what means an increase
in the particle creation due to an enhancement of the transparency.

In some situations, the $\delta-\delta^{\prime}$ and pure $\delta$ mirrors can exhibit the same total number of created particles [see, for instance, the points $C$ ($\delta$ case) and $D$ ($\delta-\delta^{\prime}$ case) belonging to the same 
level curve in Fig. \ref{fig:Total-number}].
In other situations the $\delta-\delta^{\prime}$ case exhibits a greater total number of created particles if compared with the $\delta$ case [see, for instance, the points $C$ ($\delta$ case) and $E$ ($\delta-\delta^{\prime}$ case) in Fig. \ref{fig:Total-number}].
In contrast, if we consider the $\delta$ case represented by the point $F$ and the $\delta-\delta^{\prime}$ case represented by $D$ in Fig. \ref{fig:Total-number}, the $\delta$ case exhibits a greater number of created particles if compared to the $\delta-\delta^{\prime}$ case.

It is noteworthy that the $\delta-\delta^{\prime}$ mirror ($\lambda\neq 0$), performing a spatially symmetric oscillatory motion, produces particles in asymmetric manner in both sides of the mirror [see  Eqs. (\ref{eq:spectrum-m-m}) and (\ref{eq:lambda-m-m}), 
and also Figs. \ref{spectrum-lambda} and \ref{spectrum-lambda-1}].
In Fig. \ref{fig:Ratio}, we exhibit the ratio $\mathcal{N}_{+}/\mathcal{N}_{-}$ as a function
of $\mu/\omega_0$.
For $\lambda>0$ (and $\mu\neq 0$) the production of particles in the right side of the mirror is always smaller than the production in the left side. For $\lambda<0$ the opposite occurs, since, as mentioned in the Sec. \ref{intro}, $\lambda\rightarrow-\lambda$ is equivalent to change the mirror properties from the left to the right and vice-versa.
For $\mu/\omega_0 \approx 1$ and $\lambda=1$ (see the valley point of the solid line in Fig. \ref{fig:Ratio}), we get a perfectly reflecting $\delta-\delta^{\prime}$ mirror imposing Robin (\ref{eq:RobinBC}) and Dirichlet (\ref{eq:Dirichlet-BC}) conditions to the field, being the particle creation in the right side, related to the Robin condition, strongly inhibited, remaining almost only particles created in the left side where the Dirichlet boundary condition (\ref{eq:Dirichlet-BC}) is considered.  
Note that $\mu/\omega_0 \approx 1$ corresponds to
the value $\gamma\omega_0 \approx 2.2$ (being $\gamma=2/\mu$ the Robin parameter) found in Refs. \cite{Mintz-Farina-MaiaNeto-Rodrigues-2006-I,Rego-Mintz-Farina-Alves-PRD-2013},
which is associated with a strong inhibition of the particle production for the Robin boundary condition.
Moreover, our results show that for partially reflecting $\delta-\delta^{\prime}$ mirrors there will always be a value of $\mu$ for which the asymmetry in the particle production is more strong, corresponding to the valleys of the curves in Fig. \ref{fig:Ratio},
being this asymmetry more pronounced in a perfectly reflecting $\delta-\delta^{\prime}$ mirror (solid line) for $\mu\approx 1$.
On the other hand, the symmetry ($\mathcal{N}_{+}/\mathcal{N}_{-}=1$) occurs for $\lambda=0$ [a pure $\delta$ mirror,
as shown in Eqs. (\ref{eq:spectrum-m-m}) and (\ref{pure-delta-lambda-m-m})] and for $\mu\rightarrow\infty$ (a perfectly reflecting $\delta-\delta^{\prime}$ case imposing the Dirichlet condition in both sides of the mirror).

%%%%%%%%%%%%%%%%%%%%%%%%%%%%%%%%%%%%%%%%%%%%%%%%%%%%%%%%%%%%%%%%%%%%%%%%%%%%%%%%%%
\section{Final Remarks\label{sec:Final-Remarks}}

We investigated the dynamical Casimir effect for a real massless scalar field in $1+1$
dimensions in the presence of a partially reflecting moving mirror
simulated by a $\delta-\delta^{\prime}$ point interaction. 
Specifically, considering a typical oscillatory
movement [Eq. (\ref{eq:movement}) in the monochromatic limit], we computed the spectral distribution (\ref{eq:spectrum-m-m}) and the total rate of created particles (\ref{eq:total-number-1}), this latter can be visualized in the $\mu\lambda$-plane  shown in Fig. \ref{fig:Total-number}. 
In this figure, along the dashed line ($\lambda=1$), it is shown the behavior of the total rate (\ref{eq:total-number-1}) for a perfect 
$\delta-\delta^{\prime}$ mirror, resultant from the sum of the particles produced in the left side of the mirror, which imposes the  
Dirichlet (\ref{eq:Dirichlet-BC}) condition to the field, and those produced in the right side, which imposes the Robin (\ref{eq:RobinBC}) condition.
These results are in agreement with those found in the literature \cite{Lambrecht-Jaekel-Reynaud-1996} (for Dirichlet) and \cite{Mintz-Farina-MaiaNeto-Rodrigues-2006-I} (for Robin), whereas all remaining information in $\mu\lambda$-plane was obtained in the present paper.
The behavior of the total rate (\ref{eq:total-number-1})
for a pure $\delta$ mirror (\ref{eq:Barton}) is shown along  the line $\lambda=0$.
In this case, the enhancement of the transparency (by reducing $\mu$)
leads to a monotonic reduction of the particle creation rate. 
For $\lambda\neq 0$, the $\mu\lambda$-plane exhibits
the behavior of the particle creation rate for $\delta-\delta^{\prime}$ mirrors. 
A remarkable difference between pure $\delta$ and $\delta-\delta^{\prime}$ models is that, in the latter, 
the more complex relation between phase and transparency enables
an oscillating partially reflecting
mirror to produce, via dynamical Casimir effect, a larger number of particles in comparison with
a perfect one (as illustrated by the points $A$ and $B$ in Fig. \ref{fig:Total-number}). 
In other words, the maximum coupling (in the sense that $|r_{\pm}(\omega)|=1$) between a $\delta-\delta^{\prime}$ mirror and the field does not necessarily lead to a maximum particle production, since this latter is also affected by the phases.
Furthermore, differently from the case of a typical pure $\delta$ mirror, the  $\delta-\delta^{\prime}$ mirror, performing a spatially symmetric oscillatory motion, produces particles in asymmetric manner in both sides of the mirror. A noticeable situation where almost all particles are produced in just one side of the mirror occurs for $\mu/\omega_0\approx 1$ and $\lambda=\pm 1$ (the valley of the solid curve in Fig. \ref{fig:Ratio}).
 
From our results for the massless scalar field in $1+1$ dimensions, we can infer some expected results for the dynamical Casimir effect if a $\delta-\delta^{\prime}$ mirror is considered for the massless scalar field in $3+1$ dimensions.
It is known from the literature that in $3+1$ dimensions the particle production with a Neumann condition is eleven times greater than that obtained by the Dirichlet case (see Fig. 3 in Ref. \cite{Rego-Mintz-Farina-Alves-PRD-2013}).
Since a $\delta-\delta^{\prime}$ mirror in $3+1$ dimensions with, for instance, $\mu=0$ and $\lambda= 1$, must recover a perfectly reflecting mirror imposing Neumann (right side) and Dirichlet (left side) conditions to the field, we expect that, differently from the case in $1+1$ dimensions, for a 
partially reflecting $\delta-\delta^{\prime}$ mirror with $\lambda>0$ the production of particles in the right side of the mirror can be grater than the production in the left side. 

Finally, we can infer some expected results 
if a $\delta-\delta^{\prime}$ mirror, instead of a $\delta$, is considered 
to enlarge the Maxwell Lagrangian, as discussed in Ref. \cite{Barone-2014}.
In this context, the perfect pure $\delta$ mirror is equivalent to a perfectly conducting plate \cite{Barone-2014} and, therefore, the transverse electric (TE) and transverse magnetic (TM) modes of the field obey the Dirichlet and Neumann boundary conditions respectively \cite{Maia-Neto-1996}.
Moreover, in the dynamical Casimir effect for the electromagnetic field with a perfectly conducting plate, the TM mode produces eleven times more particles than the TE mode (see Eq. (55) in Ref. \cite{Maia-Neto-1996}).
In this way, we expect that a partially reflecting $\delta$ mirror \cite{Barone-2014} also leads to asymmetric boundary conditions for the TE and TM modes on each side of the mirror, but with the TE mode in the right side of a moving mirror associated with the same particle production 
of the TE mode in the left side (the same symmetry occurring for the TM mode). 
For the case of a partially reflecting $\delta-\delta^{\prime}$ mirror, we expect that it also leads to asymmetric boundary conditions for the TE and TM modes on each side of the mirror, but now with the TE mode in the right side associated with a different particle production 
if compared with the TE mode in the left side of the moving mirror (this asymmetry also occurring for the TM mode). 
%

%%%%%%%%%%%%%%%%%%%%%%%%%%%%%%%%%%%%%%%%%%%%%%%%%%%%%%%%%%%%%%%%%%%%%%%%%%%%%%%
\begin{acknowledgments}
We thank A. L. C. Rego, B. W. Mintz, C. Farina, D. C. Pedrelli,
F. S. S. da Rosa, V. S. Alves and W. P. Pires for fruitful discussions. 
We also acknowledge the Referee for many suggestions to improve the final version of this paper.
This work was partially
supported by CAPES and CNPq Brazilian agencies.
\end{acknowledgments}

%%%%%%%%%%%%%%%%%%%%%%%%%%%%%%%%%%%%%%%%%%%%%%%%%%%%%%%%%%%%%%%%%%%%%%%%%%%%%%%
\appendix

%%%%%%%%%%%%%%%%%%%%%%%%%%%%%%%%%%%%%%%%%%%%%%%%%%%%%%%%%%%%%%%
\section{Boundary conditions for a $\delta-\delta^{\prime}$ mirror}

For the sake of completeness, here we obtain the transmission and reflection coefficients (\ref{eq:s-and-r}) associated to the Lagrangian density (\ref{eq:Munoz}) and show how they are connected to the Robin (\ref{eq:RobinBC}) and Dirichlet (\ref{eq:Dirichlet-BC}) boundary conditions.

We start reviewing some properties of the Dirac delta function and its derivative. 
Let us consider
\begin{equation}
I_1=\int_{-\infty}^{+\infty}\delta(x)f(x)g(x)\mathrm{d}x, 
\end{equation}
where $f$ is discontinuous and $g$ is continuous at $x=0$. Thus
\begin{eqnarray}
I_1 & = & \left[\int_{-\infty}^{0}+\int_{0}^{+\infty}\right] \delta(x)f(x)g(x)\mathrm{d}x 
\nonumber\qquad \\
& = & f(0^{-})g(0)\int_{-\infty}^{0}\delta(x)\mathrm{d}x
+f(0^{+})g(0)\int_{0}^{+\infty}\delta(x)\mathrm{d}x 
\nonumber\qquad \\
& = & \int_{-\infty}^{+\infty}\frac{f(0^{-})+f(0^{+})}{2}\delta(x)g(x)\mathrm{d}x.
\label{AAA0}
\end{eqnarray}
Since $g$ is arbitrary, one concludes from (\ref{AAA0}) that
\begin{equation}
\delta(x)f(x)  =  \delta(x)\frac{f(0^{+})+f(0^{-})}{2}.\label{AAAA1}
\end{equation}

There is a similar relation for the derivative of the delta function
(see, for instance, Refs. \cite{Kurasov-1996,Gadella-2009,Zolotaryuk-2010}). 
We define
\begin{equation}
I_2=\int_{-\infty}^{+\infty}\delta^{\prime}(x)f(x)g(x)\mathrm{d}x,
\label{BBB3}
\end{equation}
with $f$ and $g$, respectively, discontinuous and continuous at $x=0$.
Then, one has
\begin{eqnarray}
I_2&=&-\int_{-\infty}^{+\infty}\delta(x)[f(x)g(x)]^{\prime}\mathrm{d}x\nonumber \\
=& -&f^{\prime}(0^{-})g(0)\int_{-\infty}^{0}\delta(x)\mathrm{d}x-f^{\prime}(0^{+})g(0)\int_{0}^{+\infty}\delta(x)\mathrm{d}x\nonumber \\
&-&f(0^{-})g^{\prime}(0)\int_{-\infty}^{0}\delta(x)\mathrm{d}x-f(0^{+})g^{\prime}(0)\int_{0}^{+\infty}\delta(x)\mathrm{d}x\nonumber \\
\qquad\nonumber \\
 =& &\!\!\!\!\!\!\int_{-\infty}^{+\infty}\bigg[-\frac{f^{\prime}(0^{-})+f^{\prime}(0^{+})}{2}\delta(x)\nonumber \\
&&\qquad\left.+\frac{f(0^{-})+f(0^{+})}{2}\delta^{\prime}(x)\right]g(x)\mathrm{d}x.
\end{eqnarray}
Therefore, since $g$ is arbitrary, one concludes
\begin{equation}
\delta^{\prime}(x)f(x)=\frac{f(0^{+})+f(0^{-})}{2}\delta^{\prime}(x)-\frac{f^{\prime}(0^{+})+f^{\prime}(0^{-})}{2}\delta(x).\label{AAA3}
\end{equation}
If $f$ is continuous, Eqs. (\ref{BBB3}) and (\ref{AAA3}) takes, respectively, the simpler forms
\begin{equation}
\delta(x)f(x)  =  \delta(x)f(0),\label{AAAA1-simples}
\end{equation}
\begin{equation}
\delta^{\prime}(x)f(x) = f(0)\delta^{\prime}(x)-f^{\prime}(0)\delta(x).\label{AAA3-simples}
\end{equation}
Next, we will apply Eqs. (\ref{BBB3}) and (\ref{AAA3}) to obtain the matching conditions.

The field equation for the Lagrangian density (\ref{eq:Munoz}), in the Fourier domain, is given by 
\begin{equation}
[-\partial_{x}^{2}+2\mu\delta(x)+2\lambda\delta^{\prime}(x)]\tilde{\phi}(\omega,x)=\omega^{2}\tilde{\phi}(\omega,x).
\label{MovEq}
\end{equation}
Noticing that the field and its spatial derivative are not considered, \textit{a priori}, to be continuous at $x=0$, we shall use Eqs. (\ref{AAAA1}) and (\ref{AAA3}) rewritten as
\begin{eqnarray}
\delta(x)\tilde{\phi}(\omega,x) & = & \frac{\tilde{\phi}(\omega,0^{+})+\tilde{\phi}(\omega,0^{-})}{2}\delta(x),\label{AA1}\\
\delta^{\prime}(x)\tilde{\phi}(\omega,x) & = & \frac{\tilde{\phi}(\omega,0^{+})+\tilde{\phi}(\omega,0^{-})}{2}\delta^{\prime}(x)\nonumber \\
 &  & -\frac{\partial_{x}\tilde{\phi}(\omega,0^{+})+\partial_{x}\tilde{\phi}(\omega,0^{-})}{2}\delta(x).\qquad \label{AA2}
\end{eqnarray}
Substituting Eqs. (\ref{AA1}) and (\ref{AA2}) in (\ref{MovEq}) and integrating across $x=0$ once, we obtain
\begin{eqnarray}
-\partial_{x}\tilde{\phi}(\omega,0^{+})+\partial_{x}\tilde{\phi}(\omega,0^{-})+\mu\left[\tilde{\phi}(\omega,0^{+})+\tilde{\phi}(\omega,0^{-})\right]\nonumber \\
-\lambda\left[\partial_{x}\tilde{\phi}(\omega,0^{+})+\partial_{x}\tilde{\phi}(\omega,0^{-})\right] & = & 0.\nonumber\\
& & \label{AA0}
\end{eqnarray}
Now, integrating Eq. (\ref{MovEq}) twice [also considering Eqs. (\ref{AA1}) and (\ref{AA2})], the first one from $-L<0$ to $x$ (see, for instance, Ref. \cite{Kurasov-1993}) resulting in
\begin{eqnarray}
-\partial_{x}\tilde{\phi}(\omega,x)+\partial_{x}\tilde{\phi}(\omega,-L)+\mu\left[\tilde{\phi}(\omega,0^{+})+\tilde{\phi}(\omega,0^{-})\right]\Theta(x)\nonumber\\
-\lambda\left[\partial_{x}\tilde{\phi}(\omega,0^{+})+\partial_{x}\tilde{\phi}(\omega,0^{-})\right]\Theta(x)\nonumber\\
+\lambda\left[\tilde{\phi}(\omega,0^{+})+\tilde{\phi}(\omega,0^{-})\right]\delta(x)\nonumber\\
=\omega^{2}\int_{-L}^{x}\tilde{\phi}(\omega,x)\;\mathrm{d}x,\nonumber\\
\label{AA-intermed}
\end{eqnarray}
and integrating across $x=0$ we obtain
\begin{equation}
-\tilde{\phi}(\omega,0^{+})+\tilde{\phi}(\omega,0^{-})+\lambda[\tilde{\phi}(\omega,0^{+})+\tilde{\phi}(\omega,0^{-})]=0.
\label{AA3}
\end{equation}
Manipulating Eq. (\ref{AA3}) and substituting into Eq. (\ref{AA0}), one concludes that the field and its spatial derivative are both discontinuous at $x=0$ for $\lambda \neq 0$, and the following matching conditions are established
\begin{equation}
\tilde{\phi}(\omega,0^{+})=\frac{1+\lambda}{1-\lambda}\tilde{\phi}(\omega,0^{-}),
\label{eq:MC1}
\end{equation}
\begin{equation}
\partial_{x}\tilde{\phi}(\omega,0^{+})=\frac{1-\lambda}{1+\lambda}\partial_{x}\tilde{\phi}(\omega,0^{-})+\frac{2\mu}{1-\lambda^{2}}\tilde{\phi}(\omega,0^{-}).
\label{eq:MC2}
\end{equation}

Taking into account Eqs. (\ref{phi-00})-(\ref{eq:A10}) and (\ref{eq:matriz-espalhamento}),
the field can be written as
\begin{equation}
\phi(t,x) = \sum_{j=L,R}\int_{0}^{\infty}\mathrm{d}\omega\left[a_{j}(\omega)\Psi_{j}(\omega,x)\mathrm{e}^{-i\omega t}+H.c.\right],
\end{equation}
where
\begin{eqnarray}
\Psi_{R}(\omega,x)&=&\frac{1}{\sqrt{4\pi\omega}}\left\{ \Theta(x)\left[r_{+}(\omega)\mathrm{e}^{i\omega x}+\mathrm{e}^{-i\omega x}\right]
\right.
\cr\cr
&&\left.+\Theta(-x)s_{-}(\omega)\mathrm{e}^{-i\omega x}\right\}, 
\label{psi-R}
\end{eqnarray}
\begin{eqnarray}
\Psi_{L}(\omega,x)&=&\frac{1}{\sqrt{4\pi\omega}}\left\{ \Theta(-x)\left[r_{-}(\omega)\mathrm{e}^{-i\omega x}+\mathrm{e}^{i\omega x}\right]
\right.
\cr\cr
&&\left.+\Theta(x)s_{+}(\omega)\mathrm{e}^{i\omega x}\right\}
\label{psi-L}
\end{eqnarray}
are the left- and right-incident solutions (see, for instance, Ref. \cite{Barton-Calogeracos-1995-I}).
Equations (\ref{eq:MC1}), (\ref{eq:MC2}), (\ref{psi-R}) and (\ref{psi-L}) lead straightforwardly
to the transmission and reflection coefficients shown in Eq. (\ref{eq:s-and-r}).

The matching conditions (\ref{eq:MC1}) and (\ref{eq:MC2}) can be conveniently rewritten in the form
\begin{equation}
\varPhi_{+}=\mathbf{U}(\lambda)\varPhi_{-},
\label{CondC}
\end{equation}
where
\begin{equation}
\varPhi_{\pm}=\left(\begin{array}{c}
\mu\tilde{\phi}(\omega,0^{+})\pm i\partial_{x}\tilde{\phi}(\omega,0^{+})\\
\mu\tilde{\phi}(\omega,0^{-})\mp i\partial_x\tilde{\phi}(\omega,0^{-})
\end{array}\right),
\end{equation}
and
\begin{equation}
\mathbf{U}(\lambda)=\frac{1}{1+\lambda{}^{2}-i}\left(\begin{array}{cc}
2\lambda+i & 1-\lambda^2\\
1-\lambda^2 & -2\lambda+i
\end{array}\right).
\end{equation}
In the case $\lambda=1$ we get
\begin{equation}
\tilde{\phi}(\omega,0^{+})-(2/\mu)\partial_{x}\tilde{\phi}(\omega,0^{+})=0,
\label{c1}
\end{equation}
and
\begin{equation}
\tilde{\phi}(\omega,0^{-})=0.
\label{c2}
\end{equation}
Therefore, by taking the inverse Fourier transform of Eqs. (\ref{c1}) and (\ref{c2}), one obtains the Robin (\ref{eq:RobinBC}) and Dirichlet (\ref{eq:Dirichlet-BC}) boundary conditions respectively. Particularly, for $\mu\rightarrow 0$ in Eq. (\ref{eq:RobinBC}) the Neumann boundary condition is obtained. In addition, by taking $\lambda=0$ and $\mu\rightarrow\infty$ in Eqs. (\ref{eq:MC1}) and (\ref{eq:MC2}) and performing the Fourier inverse transform, one obtains the Dirichlet boundary condition on both sides for a pure $\delta$ mirror as shown in Eq. (\ref{limite-dirichlet-barton}).

%%%%%%%%%%%%%%%%%%%%%%%%%%%%%%%%%%%%%%%%%%%%%%%%%%%%%%%%%%%%%%%%%%%%%%%%%%%%%%%


\begin{thebibliography}{10}

\bibitem{Moore-1970}G. T. Moore, J. Math. Phys. \textbf{11}, 2679
(1970).

\bibitem{DeWitt-1975}B. S. DeWitt, Phys. Rep. \textbf{19}, 295 (1975).

\bibitem{Fulling-Davies-1976-1977}S. A. Fulling and P. C. W. Davies,
Proc. Roy. Soc. A \textbf{348}, 393 (1976);

\bibitem{Davies-Fulling-1977}P. C. W. Davies and S. A. Fulling, Proc.
R. Soc. A \textbf{356}, 237 (1977).

\bibitem{Candelas-1977}P. Candelas and D. Deutsch, Proc. R. Soc.
A \textbf{354}, 79 (1977).

\bibitem{Dodonov-2009-2010}V. V. Dodonov, J. Phys. Conf. Ser. \textbf{161},
012027 (2009); Phys. Scr. \textbf{82}, 038105 (2010).

\bibitem{Dalvit-MaiaNeto-Mazzitelli-2011}D. A. R. Dalvit, P. A. Maia
Neto, and F. D. Mazzitelli, in \emph{Casimir Physics}, edited by D.
A. R. Dalvit, P. Milonni, D. Roberts, and F. da Rosa, Lecture Notes
in Physics, Vol. 834 (Springer, New York, 2011).

\bibitem{Johansson-Nature-2011}C. M. Wilson, G. Johansson, A. Pourkabirian,
M. Simoen, J. R. Johansson, T. Duty, F. Nori, and P. Delsing, Nature
(London) \textbf{479}, 376 (2011).

\bibitem{Johansson-2009}J. R. Johansson, G. Johansson, C.M. Wilson,
and F. Nori, Phys. Rev. Lett. \textbf{103}, 147003 (2009); Phys. Rev.
A \textbf{82}, 052509 (2010).

\bibitem{Lahteenmaki-2013}P. Lähteenmäki, G. S. Paraoanu, J. Hassel,
and P. J. Hakonen, Proc. Natl. Acad. Sci. U.S.A. \textbf{110}, 4234
(2013).

\bibitem{Proposals-observation-DCE}C. Braggio, G. Bressi, G. Carugno,
C. Del Noce, G. Galeazzi, A. Lombardi, A. Palmieri, G. Ruoso, and
D. Zanello, Europhys. Lett. \textbf{70}, 754 (2005); A. Agnesi, C.
Braggio, G. Bressi, G. Carugno, G. Galeazzi, F. Pirzio, G. Reali,
G. Ruoso, and D. Zanello, J. Phys. A \textbf{41}, 164024 (2008); A.
Agnesi, C. Braggio, G. Bressi, G. Carugno, F. Della Valle, G. Galeazzi,
G. Messineo, F. Pirzio, G. Reali, G. Ruoso, D. Scarpa, and D. Zanello,
J. Phys.: Conf. Series \textbf{161}, 012028 (2009); F. X. Dezael and
A. Lambrecht, Eur. Phys. Lett. \textbf{89}, 14001 (2010); T. Kawakubo
and K. Yamamoto, Phys. Rev. A \textbf{83}, 013819 (2011); D. Faccio
and I. Carusotto, Eur. Phys. Lett \textbf{96}, 24006 (2011).

\bibitem{Ford-Vilenkin-1982}L. H. Ford and A. Vilenkin, Phys. Rev.
D \textbf{25}, 2569 (1982).

\bibitem{Haro-Elizalde-2006}J. Haro and E. Elizalde, Phys. Rev. Lett.
\textbf{97}, 130401 (2006); Phys. Rev. D \textbf{76}, 065001 (2007).

\bibitem{Eberlein-1993}G. Barton and C. Eberlein, Ann. Phys. \textbf{227},
222 (1993).

\bibitem{Jaekel-Reynaud-1992}M. T. Jaekel and S. Reynaud, Quantum
Opt. \textbf{4}, 39 (1992).

\bibitem{Lambrecht-Jaekel-Reynaud-1996}A. Lambrecht, M. T. Jaekel,
and S. Reynaud, Phys. Rev. Lett. \textbf{77}, 615 (1996).

\bibitem{Lambrecht-Jaekel-Reynaud-1998}A. Lambrecht, M. T. Jaekel,
and S. Reynaud, Eur. Phys. J. D \textbf{3}, 95 (1998).

\bibitem{Obadia-Parentani-2001}N. Obadia and R. Parentani, Phys.
Rev. D \textbf{64}, 044019 (2001); J. Haro and E. Elizalde, Phys. Rev.
D \textbf{77}, 045011 (2008); \textbf{81}, 128701 (2010).

\bibitem{Barton-Calogeracos-1995-I}G. Barton and A. Calogeracos,
Ann. Phys. \textbf{238}, 227 (1995); A. Calogeracos and G. Barton,
Ann. Phys. \textbf{238}, 268 (1995).

\bibitem{Nicolaevici-2001}N. Nicolaevici, Class. Quantum Grav. \textbf{18},
619 (2001).

\bibitem{Nicolaevici-2009}N. Nicolaevici, Phys. Rev. D \textbf{80},
125003 (2009).

\bibitem{Castaneda-Guilarte-2013} J. M. Mu\~{n}oz Casta\~{n}eda, J. M. Guilarte,
and A. M. Mosquera, Phys. Rev. D \textbf{87}, 105020 (2013).

\bibitem{MaiaNeto-Dalvit-2000}D. A. R. Dalvit and P. A. Maia Neto,
Phys. Rev. Lett. \textbf{84}, 798 (2000); P. A. Maia Neto and D. A.
R. Dalvit, Phys. Rev. A \textbf{62}, 042103 (2000).

\bibitem{Barone-2014}F. A. Barone and F. E. Barone, Phys. Rev. D
\textbf{89}, 065020 (2014).

\bibitem{Barone-2014-2}F. A. Barone and F. E. Barone, Eur. Phys. J. C \textbf{74}, 3113 (2014).

\bibitem{Parashar-Milton-Shajesh-Shaden-2012}P. Parashar, K. A. Milton, K. V. Shajesh, and M. Shaden, Phys.
Rev. D \textbf{86}, 085021 (2012).

\bibitem{Castaneda-Guilarte-2015}J. M. Mu\~{n}oz-Casta\~{n}eda and J. Mateos
Guilarte, Phys. Rev. D \textbf{91}, 025028 (2015).

\bibitem{Kurasov-1996} P. B. Kurasov, A. Scrinzi, and N. Elander, Phys. Rev. A \textbf{49}, 5095 (1994); P. Kurasov, J. Math. Anal. Appl. \textbf{201}, 297 (1996).

\bibitem{Gadella-2009} M. Gadella, J. Negro, and L. M. Nieto, Phys.
Lett. A \textbf{373}, 1310 (2009).

\bibitem{Mintz-Farina-MaiaNeto-Rodrigues-2006-I}B. Mintz, C. Farina,
P. A. Maia Neto, and R. B. Rodrigues, J. Phys. A: Math. Gen. \textbf{39},
11325 (2006); \textbf{39}, 6559 (2006).

\bibitem{Jaekel-Reynaud-1991}M. T. Jaekel and S. Reynaud, J. Phys.
I France \textbf{1}, 1395 (1991).

\bibitem{MIT}M. F. Maghrebi, R. Golestanian, and M. Kardar, Phys.
Rev. D \textbf{87}, 025016 (2013); see also references therein.

\bibitem{Moysez}H. M. Nussenzveig, \emph{Causality and dispersion
relations} (Academic Press, New York, 1972).

\bibitem{Silva-Braga-Rego-Alves-2015} J. D. Lima Silva, A. N. Braga,
A. L. C. Rego, and D. T. Alves, Phys. Rev. D \textbf{92}, 025040 (2015).

\bibitem{Rego-Mintz-Farina-Alves-PRD-2013} A. L. C. Rego, B. W. Mintz,
C. Farina, and D. T. Alves, Phys. Rev. D \textbf{87}, 045024 (2013).

\bibitem{Silva-Farina-2011}H. O. Silva and C. Farina, Phys. Rev.
D \textbf{84}, 045003 (2011).

\bibitem{Maia-Neto-1996}P. A. Maia Neto and L. A. S. Machado, Phys. Rev.
A \textbf{54}, 3420 (1996).

\bibitem{Casimir-1948} H. B. G. Casimir, Proc. K. Ned. Akad. Wet.
\textbf{51}, 793 (1948).

\bibitem{Kurasov-1993} P. B. Kurasov and N. Elander, Preprint MSI 93-7, ISSN-1100-214X, Stockholm, Sweden (1993).

\bibitem{Zolotaryuk-2010} A. V. Zolotaryuk, Phys. Lett. A \textbf{374}, 1636 (2010).

\end{thebibliography}
\end{document}